\documentclass[prd,showpacs,amsmath,amssymb,nofootinbib,floatfix]{revtex4}
\usepackage{graphicx,color,dcolumn,booktabs}

\begin{document}
\title{Re-Study on the wave functions of $\Upsilon(nS)$ states in LFQM and the radiative decays of $\Upsilon(nS)\to \eta_b+\gamma$}

\author{Hong-Wei Ke}\email{khw020056@hotmail.com}\affiliation{School of Science, Tianjin University, Tianjin 300072, China}

\author{Xue-Qian Li}\email{lixq@nankai.edu.cn}\affiliation{School of Physics, Nankai University, Tianjin 300071, China}
\author{Zheng-Tao Wei}\email{weizt@nankai.edu.cn}\affiliation{School of Physics, Nankai University, Tianjin 300071, China}
\author{Xiang Liu$^{1,2}$\footnote{Corresponding author}}\email{xiangliu@lzu.edu.cn}\affiliation{
$^1$Research Center for Hadron and CSR Physics,
Lanzhou University $\&$ Institute of Modern Physics of CAS, Lanzhou 730000, China\\
$^2$School of Physical Science and Technology, Lanzhou University, Lanzhou 730000, China}

\date{\today}
\begin{abstract}
\noindent The Light-front quark model (LFQM) has been applied to
calculate the transition matrix elements of heavy hadron decays.
However, it is noted that using the traditional wave functions of
the LFQM given in literature, the theoretically determined decay
constants of the $\Upsilon(nS)$ obviously contradict to the data. It
implies that the wave functions must be modified. Keeping the
orthogonality among the $nS$ states and fitting their decay
constants we obtain a series of the wave functions for
$\Upsilon(nS)$. Based on these wave functions and by analogy to the
hydrogen atom, we suggest a modified analytical form for the
$\Upsilon(nS)$ wave functions. By use of the modified wave
functions, the obtained decay constants are close to the
experimental data.  Then we calculate the rates of radiative decays
of $\Upsilon(nS)\to \eta_b+\gamma$. Our predictions are consistent
with the experimental data on decays $\Upsilon(3S)\to \eta_b+\gamma$
within the theoretical and experimental errors.

\end{abstract}

\maketitle

\section{introduction}

Since the relativistic and higher-order $\alpha_s$ corrections are
less important for bottomonia than for any other $q\bar q$ systems,
study on bottomonia may offer more direct information about the
hadron configuration and application of the perturbative QCD. The
key problem is how to deal with the hadronic transition matrix
elements which are fully governed by the non-perturbative QCD
effects. Many phenomenological models have been constructed and
applied. Each of them has achieved relative successes, but since
none of them are based on any well established underlying theories,
their model parameters must be obtained by fitting data. By doing
so, some drawbacks of the model are exposed when applying to deal
with different phenomenological processes. Thus one needs to
continuously modify the model or re-fit its parameter, if not
completely negate it. The light front quark model is one of such
models. It has been applied to calculate the hadronic transitions
and generally considered as a successful one. The model contains a
Gaussian-type wavefunction whose parameters should be determined in
a certain way.

The Gaussian-type wavefunction was recommended by the authors of
Refs. \cite{Jaus:1999zv,Cheng:2003sm} and most frequently the
wavefunction for harmonic oscillator is adopted which we refer as
the traditional LFQM wavefunction. As we employed the traditional
LFQM wave functions to calculate the branching ratios of
$\Upsilon(nS)\rightarrow\eta_b+\gamma$, some obvious
contradictions between the theoretical predictions and
experimental data emerged. Namely, the predicted $
\mathcal{B}(\Upsilon(2S)\rightarrow\eta_b+\gamma)$ was one order
larger than the experimental upper bound \cite{Ke:2010tk}.
Moreover, as one carefully investigates the wave functions, he
would face a serious problem. If the traditional wave functions
were employed, the decay constants of $\Upsilon(nS)$ ($f_V$) would
increase for higher $n$. It obviously contradicts to the
experimental data and the physics picture which tells us that the
decay constant of a $nS$ state is proportional to its wavefunction
at origin which manifests the probability that the two
constituents spatially merge, so for excited states the
probability should decrease. Thus the decay constants should be
smaller as $n$ is larger. The experimental data confirm this trend.
But the theoretical calculations with the traditional
wave functions result in an inverse order. To overcome these
problems, one may adopt different model parameters (refers to $\beta$)
by fitting individual $n$'s decay constants as done in \cite{Ke:2010tk,Wang:2010np},
but the orthogonality among the $nS$ states is broken. In this work, we try
to modify the harmonic oscillator functions and introduce an explicit
$n$-dependent form for the wave functions. Keeping the orthogonality
among the $nS$ states ($n=1,...5$), we modify the LFQM wave functions. By
fitting the decay constants of $\Upsilon(nS)$, the concerned model
parameters are fixed. Besides fitting the decay constants of the $\Upsilon(nS)$ family,
one should test the applicability of the model in other processes.
We choose the radiative decays of $\Upsilon(nS)\rightarrow
\eta_b+\gamma$ as the probe. As a matter of fact, those radiative
decays are of great significance for understanding the hadronic
structure of bottomonia family.

Indeed, the spin-triplet state of bottomonia $\Upsilon(nS)$ and the
P-states $\chi_b(nP)$ were discovered decades ago, however the
singlet state $\eta_b$ evaded detection for a long time, even though
much efforts were made. Many phenomenological researches on $\eta_b$
have been done by some groups
\cite{Hao:2007rb,Ebert:2002pp,Motyka:1997di,Liao:2001yh,Recksiegel:2003fm,
Gray:2005ur,Eichten:1994gt,Ke:2007ih}. Different theoretical
approaches result in different level splitting $\Delta
M=\Upsilon(1S)-\eta_b(1S)$.  In \cite{Recksiegel:2003fm} the
authors used an improved perturbative QCD approach to get $\Delta
M=44$ MeV; using the potential model suggested in
\cite{Buchmuller:1980su} Eichten and Quigg estimated $\Delta M=87$
MeV \cite{Eichten:1994gt}; in Ref. \cite{Motyka:1997di} the authors
selected a non-relativistic Hamiltonian with spin dependent
corrections to study the spectra of heavy quarkonia and got $\Delta
M$=57 MeV; the lattice prediction is $\Delta M$=51 MeV \cite{Liao:2001yh},
whereas the lattice result calculated in Ref. \cite{Gray:2005ur} was
$\Delta M=64\pm14$MeV. Ebert $et\, al.$ \cite{Ebert:2002pp} directly
studied spectra of heavy quarkonia in the relativistic quark model
and gave $m_{\eta_b}=9.400$ GeV. The dispersion of the values may
imply that there exist some ambiguities in our understanding about
the structures of the $b\bar b$ family.

\begin{center}
\begin{figure}[htb]
\begin{tabular}{c}
\scalebox{1.2}{\includegraphics{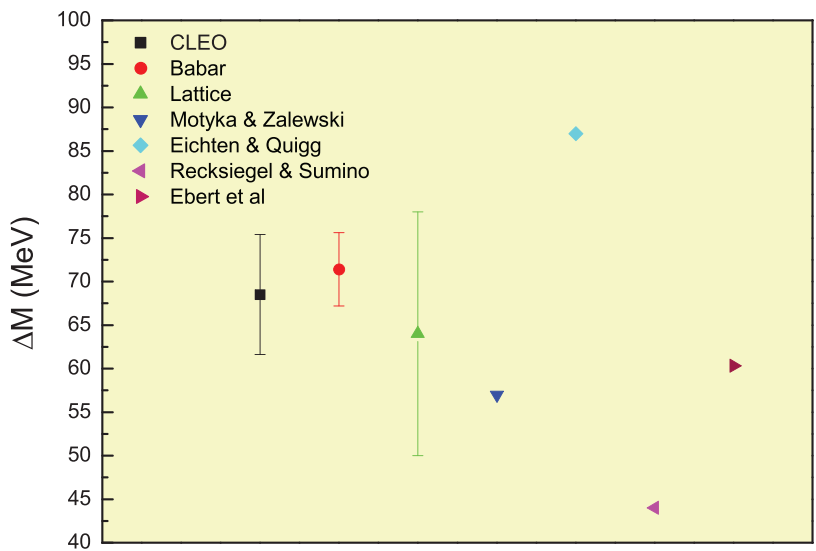}}
\end{tabular}
\caption{$\Delta M$ coming from different experimental measurement
and theoretical work.}\label{DM}
\end{figure}
\end{center}

The Babar Collaboration \cite{:2008vj} first measured
$\mathcal{B}(\Upsilon(3S)\rightarrow\gamma\eta_b)=(4.8\pm0.5\pm0.6)\times10^{-4}$,
and determined $m_{\eta_b}=9388.9^{+3.1}_{-2.3}\pm2.7$ MeV, $\Delta
M= 71.4^{+3.1}_{-2.3}\pm2.7$ MeV in 2008. New data
$m_{\eta_b}=9394.2^{+4.8}_{-4.9}\pm2.0$ MeV and
$\mathcal{B}(\Upsilon(2S)\rightarrow\gamma\eta_b)=(3.9\pm1.1^{+1.1}_{-0.9})\times10^{-4}$
were released in 2009 \cite{:2009pz}. More recently the CLEO
Collaboration \cite{Bonvicini:2009hs} confirmed the observation of
$\eta_b$ using the database of 6 million $\Upsilon(3S)$ decays and
assuming $\Gamma(\eta_b)\approx$10 MeV, they obtained
$\mathcal{B}(\Upsilon(3S)\rightarrow\gamma\eta_b)=(7.1\pm1.8\pm1.1)\times10^{-4}$,
$m_{\eta_b}=9391.8\pm6.6\pm2.0$ MeV and the hyperfine splitting
$\Delta M= 68.5\pm6.6\pm2.0$ MeV, whereas using the database with 9
million $\Upsilon(2S)$ decays they obtained
$\mathcal{B}(\Upsilon(2S)\rightarrow\gamma\eta_b)<8.4\times10^{-4}$
at 90\% confidential level. It is noted that the data of the two
collaborations are in accordance on $m_{\eta_b}$, but the central
values of $\mathcal{B}(\Upsilon(3S)\rightarrow\gamma\eta_b)$ are
different. However, if the experimental errors are taken into
account, the difference is within one standard deviation.

Some theoretical works
\cite{Radford:2009qi,Colangelo:2009pb,Seth:2009ba} are devoted to
account the experimental results. In Ref. \cite{Ebert:2002pp} the
authors studied these radiative decays and estimated
$\mathcal{B}(\Upsilon(3S)\rightarrow\eta_b+\gamma)=4\times10^{-4}$,
$\mathcal{B}(\Upsilon(2S)\rightarrow\eta_b+\gamma)=1.5\times10^{-4}$
and $\mathcal{B}(\Upsilon(1S)\rightarrow\eta_b+\gamma)=1.1\times10^{-4}$
with the mass $m_{\eta_b}$ = $9.400$ GeV. Their results about
$m_{\eta_b}$ and $\mathcal{B}(\Upsilon(3S)\rightarrow\eta_b+\gamma)$
are close to the data. The authors of Ref. \cite{Choi:2007se}
systematically investigated the magnetic dipole transition
$V\rightarrow P\gamma$ in the light-front quark model (LFQM)
\cite{Jaus:1999zv,Cheng:2003sm,Hwang:2006cua,Wei:2009nc}. In the
QCD-motivated approach there are several free parameters, i.e., the
quark mass and $\beta$ in the wave function (the notation of $\beta$
was given in the aforementioned literatures) which are fixed by the
variational principle, then
$\mathcal{B}(\Upsilon(1S)\rightarrow\eta_b+\gamma)$ was calculated
and the central value is $8.4\,({\rm or}\,7.7)\times 10^{-4}$
\footnote{The different values correspond to the different
potentials adopted in the calculations.}. It is also noted that the
mass of $m_{\eta_b}=9.657\,({\rm or} \,9.295)$ GeV presented in Ref.
\cite{Choi:2007se} deviates from the data listed before, so we are
going to re-fix the parameter $\beta$ in other ways namely we fix
the parameter $\beta$ by fitting data.

Since experimentally, $m_{\eta_b}$ is determined by
$\mathcal{B}(\Upsilon(nS)\rightarrow\eta_b+\gamma)$ and a study on
the radiative decays  can offer us much information about the
characteristics of $\eta_b$, one should carefully investigate the
transition within a relatively reliable theoretical framework. That
is the aim of the present work, namely we will evaluate the hadronic
matrix element in terms of our modified LFQM.

This paper is organized as follows. After this  introduction, in
section II we discuss how to modify the traditional wave functions
in LFQM. We present the formula to calculate the form factors for
$V\rightarrow P\gamma$ in the LFQM and  numerical results  in
section III. The section IV is devoted to our conclusion and
discussion.

\section{The modified  wave functions for the  radially excited
states }

When the LFQM is employed to calculate the decay constants and
form factors, one needs the wave functions of the concerned
hadrons. In most cases, the wave functions of harmonic oscillator
are adopted. In the works \cite{Jaus:1999zv, Cheng:2003sm,
Hwang:2006cua,Wei:2009nc,Choi:2007se,Ke:2009mn}, only the wave
function of the radially ground state is needed, but when in the
processes under consideration radially excited states of are
involved, their wave functions should also be available. In
\cite{Faiman:1968js, Isgur:1988gb}, the traditional wave functions
$\varphi$ for $1S$ and $2S$ states in configuration space from
harmonic oscillator are given as
\begin{eqnarray} \label{eq:12S}
 \varphi^{1S}(r)&=&\Big(\frac{\beta^2}{\pi}\Big)^{3/4}{\exp}\Big(-\frac{1}{2}\beta^2\mathbf{r}^2\Big),\nonumber\\
 \varphi^{2S}(r)&=&\Big(\frac{\beta^2}{\pi}\Big)^{3/4}{\exp}\Big(-\frac{1}{2}\beta^2\mathbf{r}^2\Big)\frac{1}{\sqrt{6}}
  \Big(3-2\beta^2\mathbf{r}^2\Big).
\end{eqnarray}
In order to maintain the orthogonality among $nS$ states, the
parameter $\beta$ in the above two functions are the same. The wave
functions for other $nS$ state  can be found in  Appendix A.

The decay constants of the $nS$ states are directly
proportional to the wave function at the origin
\begin{eqnarray}\label{24}
 f_V\propto \varphi(r=0).
\end{eqnarray}If we simply adopt the wave functions of harmonic
oscillator for all of them as we do for the $1S$ state, then we find
the wave functions at the origin, i.e. $\varphi(r=0)$ (see Appendix
for details) rises with increase of $n$ (the principle quantum
number) which means the decay constants would increase for larger
$n$. For example, by Eq. (\ref{eq:12S}) the ratio of wave
functions of $2S$ and $1S$ states at the origin is $3/\sqrt{6}>1$.

The decay constants $f_V$ of $\Upsilon(nS)$ are extracted from the processes
$\Gamma(\Upsilon(nS)\rightarrow e^+e^- )$ with
\begin{eqnarray}\label{25}
\Gamma(V\rightarrow
e^+e^-)=\frac{4\pi}{27}\frac{\alpha^2}{m_V}f^2_{V},
\end{eqnarray}
where $V$ represents $\Upsilon(nS)$ and $m_V$ its mass. By use of the experimental
data from PDG \cite{PDG08}, we obtain the experimental values for $f_V$ which are
listed in Table \ref{tab:decay}.  Obviously,  the decay constant becomes
smaller as $n$ is larger.

In LFQM, the formula for calculating the vector meson decay constant is given by
\cite{Jaus:1999zv,Cheng:2003sm}
\begin{eqnarray}\label{26}
f_V&=&\frac{\sqrt{N_c}}{4\pi^3M}\int dx\int
d^2k_\perp\frac{\phi(nS)}{\sqrt{2x(1-x)}\tilde M_0}
\biggl[xM_0^2-m_1(m_1-m_2)-k^2_\perp+\frac{m_1+m_2}{M_0+m_1+m_2}k^2_\perp\biggl],
\end{eqnarray}
where $m_1=m_2=m_b$ and other notations are collected in the
Appendix. In the calculation we set $m_b=5.2$ GeV following
\cite{Choi:2007se} and the decay constant of $\Upsilon(1S)$ is used
to determine the parameter $\beta_\Upsilon$ as the input. We obtain
$\beta_\Upsilon=1.257\pm$0.006 GeV corresponding to $f^{\rm
exp}_{\Upsilon(1S)}=715\pm 5$ MeV.   In order to illustrate the
dependence of our results on $m_b$, we re-set $m_b=4.8$ GeV to
repeat our calculation, then by fitting the same data, we fix
$\beta_\Upsilon=1.288\pm$0.006 GeV and all the results are clearly
shown in the following tables. The $f_\Upsilon^{\rm T}$ in Table
\ref{tab:decay} are the decay constant calculated in the traditional
wave functions. These results expose an explicit contradictory
trend. Thus, our calculation indicates that if the traditional wave
functions are used, the obtained decay constants of $\Upsilon(nS)$
would sharply contradict to the experimental data.

\begin{table}
\caption{ The decay constants of $\Upsilon(nS)$ (in the unit of
MeV). The column ``$f_\Upsilon^{\rm T}$" represents the theoretcal
predictions with the traditional wave function in LFQM. The column
``$f_\Upsilon^{\rm M}$" represents the prediction with our modified
wave function and the values in the brackets are the corresponding
values with $m_b=4.8$ GeV as input. (The other values are
corresponding to $m_b=5.2$ GeV.)} \label{tab:decay}
\begin{tabular}{c|c|c|c}\hline\hline
 ~~~~~~~nS~~~~~~~   &  ~~~~~~$f_\Upsilon^{\rm exp}$~~~~~~   &
 ~~~~~~$f_\Upsilon^{\rm T}$~~~~~~ &  ~~~~~~$f_\Upsilon^{\rm M}$~~~~~~   \\\hline
 1S    & 715$\pm$5    &  715$\pm$5       &  715$\pm 5$ (715$\pm 5$)           \\
 2S    & 497$\pm$5    &  841$\pm$7       &  497$\pm 5$ (498$\pm 5$)    \\
 3S    & 430$\pm$4    &  925 $\pm$8      &  418$\pm 5$ (419$\pm 4$)   \\
 4S    & 340$\pm$19   &  993  $\pm$8     &  378$\pm 4$ (397$\pm 4$)   \\
 5S    & 369$\pm$42   &  1040 $\pm$9    &  349$\pm 4$  (351$\pm 4$) \\\hline\hline
\end{tabular}
\end{table}

As aforementioned, the wave functions must be modified. Our strategy is to establish a
new Gaussian-type wave function which is different from that of harmonic oscillator.
As modifying the wave functions,  several principles must be respected:

(1) The wave function of $1S$  should not change because its
application for dealing with various processes has been tested and
the results indicate that it works well;

(2) The number of nodes of $nS$ should not be changed;

(3) A factor may be added into the wave functions which should
uniquely depend on $n$ in analog to the  wave function of the
hydrogen-like atoms which is written as
$R_n(r)=P^{hydr}_n(r)e^{-{Zr\over na_0}}$, where $P_n(r)$ is a
polynomial and $Z$ is the atomic number, $a_0$ is the Bhor radius;

(4) Using the new Gaussian-type wave function, the contradiction for the decay
constants can be solved.

In the LFQM, we only need the wave functions in the momentum space. Fourier
transformation gives us the corresponding forms in the momentum space, see the Appendix
for details. The $1S$ wave function is remained and used to fix the model parameter.
Now let us investigate the  wave function of $2S$. According to the
analog to the hydrogen-like atom, we introduce a factor $g_2$ represents $n$-dependence to the
exponential in the wave function of $2S$, thus the wave function of $2S$ is
changed to
\begin{eqnarray}\label{2S}
\psi_{_M}^{2S}(\mathbf{p}^2)=\Big(\frac{\pi}{\beta^2}\Big)^{3/4}{\exp}\Big(-g_2\frac{\mathbf{p}^2}{2\beta}
\Big) \Big(a+b\frac{\mathbf{p}^2}{\beta^2}\Big),
\end{eqnarray}
where the subscript $M$ denotes the modified function. Then by
requiring it to be orthogonal to that of $1S$ and normalizing the
wave function, we determine the parameters $a$ and $b$ in the
modified wave function of $2S$. With this new wave function of $2S$, we demand
the theoretical decay constant be consistent with data so $g_2$ should fall into a
range determined by the experimental errors. Going on, we obtain the modified
wave function of $3S$ and that for 4S and 5S as well. In this case
the modified wave functions of $nS$ states are more complicated
than the traditional ones.

We have gained a series of numerical $g_n$'s by the principles we
discussed above, then we wish to guess an analytical factor $g_n$
which is close to the numerical values of the series. We find that
if $g_n=n^\delta$( $\delta=1/1.82$) is set, we almost recover the
numerical series. Thus the wave function of the $nS$ state in the
momentum space can be written as
\begin{eqnarray}
\psi_{_M}^{nS}({\bf p}^2)=P_n({\bf
p}^2){\exp}\Big(-n^\delta\frac{\mathbf{p}^2}{2\beta^2} \Big),
\end{eqnarray}
where $P_n({\bf p}^2)$ is a polynomial in ${\bf p}^2$. The
corresponding wave function of the $nS$ state in the configuration
space can be written as
\begin{eqnarray}
\psi_{_M}^{nS}(r)=P'_n({\bf
r}^2){\exp}\Big(-\frac{\beta^2\mathbf{r}^2}{2n^\delta} \Big).
\end{eqnarray}
Comparing with the case of the hydrogen-like atoms which the
$nS$-wave functions are written as
\begin{eqnarray}
R_{n0}(r)=P^{\rm hydr}_n(r)\exp\Big({-Zr\over na_0}\Big).
\end{eqnarray}
in the configuration space, where $P^{hydr}_n(r)$ is a polynomial in
$r$. The factor $1/n$ in the exponential power is obtained by
solving the Schr\"odinger equation where only the Coulomb potential
exists. To modify the wave functions we get the factors numerically
for all the $nS$ states, then ``guess" its analytical form. In
the LFQM, the factor $1/n^\delta$ is introduced to fit the
experimental data for $nS$ decay constants. Definitely this
analytical form is not derived from an underlying theory, such as
that for the hydrogen atom, thus the dependence on $n$ is only an
empirical expression. But we are sure that if the model is correct
and our guess is reasonable, it should be obtained from QCD (maybe
non-perturbative QCD).  It is noted that the experimental errors
are large, so that other forms for $g_n$ might also be possible. The
theoretical estimation of the decay constants of $\Upsilon(nS)$
($f_\Upsilon^{\rm M}$) are also presented in Table \ref{tab:decay}.
The modified wave functions seem to work well and they could be used
for evaluating $\mathcal{B}(\Upsilon(nS)\rightarrow\eta_b+\gamma)$.

\section{the transition of $\Upsilon(nS)\to \eta_b +\gamma$ }

In this section, we calculate the branching ratios of
$\Upsilon(nS)\to \eta_b +\gamma$ in terms of the modified wave functions
derived in the above section.

\subsection{Formulation of $\Upsilon(nS)\to \eta_b +\gamma$ in the LFQM}

\begin{center}
\begin{figure}[htb]
\begin{tabular}{cc}
\scalebox{0.8}{\includegraphics{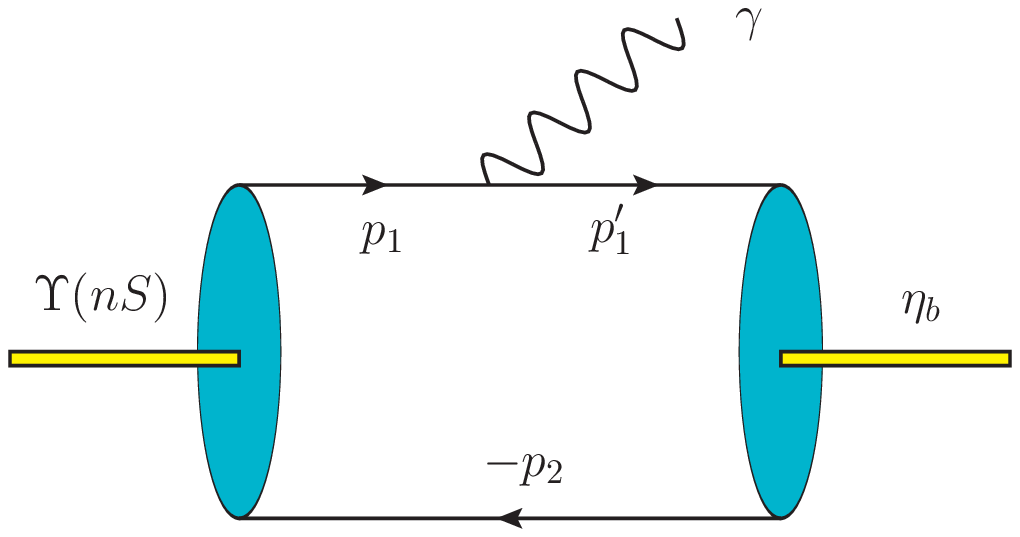}}&\raisebox{-2.5em}
{\scalebox{0.8}{\includegraphics{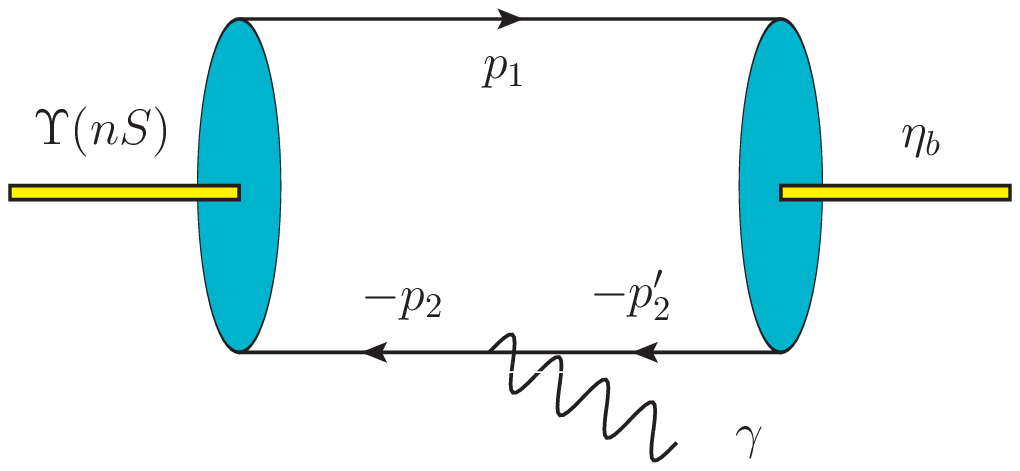}}}
\end{tabular}
\caption{Feynman diagrams depicting the radiative decay $\Upsilon(nS)\to \eta_b+\gamma$.\label{fig:LFQM}}
\end{figure}
\end{center}

The Feynman diagrams describing $\Upsilon(nS)\to \eta_b+ \gamma$ are
plotted in Fig. \ref{fig:LFQM}. The transition amplitude of
$\Upsilon(nS)\to \eta_b+ \gamma$ can be expressed in terms of the form factor
$\mathcal{F}_{\Upsilon(nS)\to\eta_b}(q^2)$ which is defined as
\cite{Choi:2007se,Hwang:2006cua}
\begin{eqnarray}\label{2S1}
&&\langle \eta_b(\mathcal{P}')|J_{em}^\mu|\Upsilon(\mathcal{P},h)\rangle
 =ie\,\varepsilon^{\mu\nu\rho\sigma}\epsilon_\nu(\mathcal{P},h)q_\rho
\mathcal{P}_\sigma\mathcal{ F}_{\Upsilon(nS) \to\eta_b}(q^2),
\end{eqnarray}
where $\mathcal{P}$  and $\mathcal{P}'$ are the four-momenta of
$\Upsilon(nS)$ and $\eta_b$. $q=\mathcal{P}-\mathcal{P}'$ is the
four-momentum of the emitted photon and
$\epsilon_\nu(\mathcal{P},h)$ denotes the polarization vector of
$\Upsilon(nS)$ with helicity $h$. For applying the LFQM, we first
let the photon be virtual, i.e. leave its mass-shell $q^2=0$ into
the un-physical region of $q^2<0$. Then $\mathcal{ F}_{\Upsilon(nS)\to
\eta_b}(q^2)$ can be obtained in the $q^+=0$ frame with $q^2=q^+q^-
- {\bf q}^2_\perp=-{\bf q}^2_\perp<0$. Then we just analytically
extrapolate $\mathcal{ F}_{\Upsilon(nS) \to\eta_b}({\bf q}^2_\perp)$
from the space-like region to the time-like region ($q^2\geq 0$). By
taking the limit $q^2\rightarrow 0$, one obtains $\mathcal{
F}_{\Upsilon(nS)\to \eta_b}( q^2=0)$.

By means of the light front quark model, one can obtain the
expression of the form factor $\mathcal{ F}_{\Upsilon(nS)\to
\eta_b}(q^2)$ \cite{Choi:2007se}:
\begin{eqnarray}\label{21}
\mathcal{ F}_{\Upsilon(nS)\to\eta_b}(q^2)= e_bI(m_1,m_2,q^2) + e_{b}
I(m_2,m_1,q^2),
\end{eqnarray}
where $e_{b}$ is the electrical charge for the bottom quark,
$m_1=m_2=m_b$ and
\begin{eqnarray}\label{22}
I(m_1,m_2,q^2) &=&\int^1_0 \frac{dx}{8\pi^3}\int d^2{\bf k}_\perp
\frac{\phi(x, {\bf k'}_\perp)\phi(x,{\bf k}_\perp)}{x_1\tilde{M_0}\tilde{M'_0}}
\times \biggl\{{\cal A} + \frac{2} {{\cal M}_0} [{\bf k}^2_\perp -
\frac{({\bf k}_\perp\cdot{\bf q}_\perp)^2}{{\bf q}^2_\perp}]\biggr\}.
\end{eqnarray}
where ${\cal A}=x_2m_1+x_1m_2$, $x=x_1$ and the other variables in Eq.
(\ref{22}) are defined in Appendix. In the covariant light-front
quark model, the authors of \cite{Hwang:2006cua} obtained the
same  form factor $\mathcal{ F}_{\Upsilon(nS)\to
\eta_b}(\mathbf{q}^2)$. The decay width for
$\Upsilon(nS)\rightarrow \eta_b+\gamma$ is easily achieved
\begin{eqnarray}\label{23}
\Gamma(\Upsilon(nS)\rightarrow
\eta_b+\gamma)=\frac{\alpha}{3}\bigg[\frac{m_{\Upsilon(nS)}^2-m_{\eta_b}^2}{2m_{\Upsilon(nS)}}\bigg]^3
\mathcal{ F}^2_{\Upsilon(nS) \to\eta_b}(0).
\end{eqnarray}
where $\alpha$ is the fine-structure constant and
$m_{\Upsilon(nS)},\; m_{\eta_b}$ are the masses of $\Upsilon(nS)$
and $\eta_b$ respectively.

\subsection{Numerical results  }

Now we begin to evaluate the transition rates of
$\Upsilon(2S)\rightarrow\eta_c+\gamma$ with the modified wave
functions. We still use the values of  $m_b=5.2$ GeV and
$\beta_\Upsilon=1.257\pm 0.006$ GeV given in last section. The
parameter $\beta_{\eta_b}$ is unknown, we determine it from
$\Upsilon(2S)\rightarrow\gamma\eta_b$ process. Comparing with the
data $\mathcal{B}(\Upsilon(2S)\rightarrow\gamma\eta_b)=3.9\times
10^{-4}$ \cite{:2009pz}, we obtain $\beta_{\eta_b}=1.246\pm0.005$
GeV which is consistent with our expectation, namely it is close to
the value of $\beta_{\Upsilon}=1.257$ GeV. Under the heavy quark
limit, they should be exactly equal, and the deviation must be of
order $\mathcal{O}(1/m_b)$ which is small \cite{Isgur:1989vq}. With
these parameters, we can calculate the branching ratios
$\mathcal{B}(\Upsilon(1S)\rightarrow\eta_b+\gamma)$,
$\mathcal{B}(\Upsilon(3S)\rightarrow\eta_b+\gamma)$,
$\mathcal{B}(\Upsilon(4S)\rightarrow\eta_b+\gamma)$, and
$\mathcal{B}(\Upsilon(5S)\rightarrow\eta_b+\gamma)$.  The numerical
results are presented in the column ``$\mathcal{B}_{\rm I}^{\rm M}$"
of Table \ref{tab:etab2}. Indeed, the b-quark mass is an uncertain
parameter which cannot be directly measured and in some literatures,
different values for b-quark mass have been adopted. To see how
sensitive to the b-quark mass the result would be,  we also present
the numerical results with $m_b=4.8$ GeV, $\beta_\Upsilon=1.288\pm
0.006$ GeV and $\beta_{\eta_b}=1.287\pm0.005$ GeV in the column
``$\mathcal{B}_{\rm II}^{\rm M}$" of Table \ref{tab:etab2}.  The
results in the column ``$\mathcal{B}^{\rm T}$" of Table
\ref{tab:etab2} are obtained with the traditional wave functions.
Apparently, as the modified wave functions  are employed, the
theoretical  predictions  on branching ratios of the radiative
decays are much improved, namely deviations from the data is
diminished. About the numerical results, some comments are given as
following:

(1)  Comparing the results shown in column $\mathcal{B}_{\rm I}^{\rm
M}$ with those in column $\mathcal{B}_{\rm II}^{\rm M}$, we can find
that they are not sensitive to $m_b$.

(2) For the decay $\Upsilon(1S)\rightarrow\eta_b+\gamma$, our
prediction on the branching ratio is about $2.0\times 10^{-4}$.
This mode should be observed soon in the coming experiment. Our
prediction is consistent with the results of Ref.
\cite{Ebert:2002pp,Choi:2007se}. The branching ratio is not
sensitive to  $\beta_{\eta_b}$, but sensitive to the mass
splitting $\Delta M$. That is easy to understand. Since the decay
width is proportional to $(\Delta M)^3$, thus as $\Delta M$ is
small, i.e., the masses of initial and daughter mesons are close
to each other, any small change of $m_{\eta_b}$ can lead to a
remarkable difference in the theoretical prediction on the
branching ratio.
Thus the accurate measurement on
$\mathcal{B}(\Upsilon(1S)\rightarrow\eta_b+\gamma)$ will be a great
help to determine the mass of $m_{\eta_b}$.

(3) The process of $\Upsilon(2S)\rightarrow\eta_b+\gamma$ is used
as an input to determine the parameter of $\eta_b$. The prediction
of $\Upsilon(3S)\rightarrow\eta_b+\gamma$ is in accordance with
the experimental data by the order of magnitude. After taking into
account the experimental and theoretical errors, they can be
consistent. This result could be of relatively large errors,
because we only use four parameters ($m_b$, $\beta_\Upsilon$,
$\beta_{\eta_b}$, $\alpha$) to determine five decay constants and
three branching ratios for $\Upsilon(1S,2S,3S)\to \eta_b+\gamma$
and all of them possess certain errors.

(4) The branching ratios for the processes
$\Upsilon(4S)\rightarrow\eta_b+\gamma$ and
$\Upsilon(5S)\rightarrow\eta_b+\gamma$ are at the order of
$10^{-8}$, it is nearly impossible to be observed  in the near
future if there aren't other mechanisms to enhance them.

(5) As an application, we predict the decay constant of $\eta_b$ in
terms of the model parameters we obtained above. We calculate the
branching ratio of
$\mathcal{B}$$(\Upsilon(2S)\rightarrow\gamma\eta_b)$ in the LFQM. By
fitting data we fix the concerned model parameters for $\eta_b$, and
then with them we predict the decay constant of  $\eta_b$ in the
same framework of the LFQM \cite{Jaus:1999zv,Cheng:2003sm}. In the
calculations, b-quark mass $m_b$ and $\beta_{\eta_b} $ are input
parameters.

To show how sensitive the results are to the parameters, we use the
two sets of input parameters given above, and the corresponding
results are as follows: $f_{\eta_b}=567$ MeV when $m_b=5.2$GeV and
$\beta_{\eta_b}=1.246$GeV ; $f_{\eta_b}=604$ MeV when $m_b=4.8$GeV
and $\beta_{\eta_b}=1.287$GeV. For a comparison, we deliberately
change only $m_b$ while keeping $\beta_{\eta_b}$ unchanged to repeat
the calculation and obtain $f_{\eta_b}=591$ MeV when $m_b=5.2$GeV
and $\beta_{\eta_b}=1.287$GeV. It is noted that $f_{\eta_b}$ is more
sensitive to $\beta_{\eta_b}$, rather than $m_b$.

\begin{table}
\caption{ The branching ratios of
$\Upsilon(nS)\rightarrow\gamma\eta_b$. In the column
``$\mathcal{B}_{\rm I}^{\rm M}$", $m_b=5.2$GeV,
$\beta_\Upsilon=1.257\pm 0.006$ GeV and $\beta_{\eta_b}=1.246\pm
0.005$ GeV. In the column ``$\mathcal{B}_{\rm II}^{\rm M}$",
$m_b=4.8$GeV, $\beta_\Upsilon=1.288\pm 0.006$ GeV and
$\beta_{\eta_b}=1.287\pm0.005$ GeV. In the column
``$\mathcal{B}^{\rm T}$", $m_b=5.2$GeV, $\beta_\Upsilon=1.257\pm
0.006$ GeV and $\beta_{\eta_b}=1.249\pm0.005$ GeV.}
\begin{ruledtabular}
\begin{tabular}{ccccc}
 Decay mode & $\mathcal{B}_{\rm I}^{\rm M}$ & $\mathcal{B}_{\rm II}^{\rm M}$ & $\mathcal{B}^{\rm T}$& Experiment  \\\hline
 $\Upsilon(1S)\rightarrow \eta_b+\gamma$ & $(1.94\pm 0.41)\times 10^{-4}$ &
  $(2.24\pm 0.47)\times 10^{-4}$  & $(1.94\pm 0.42)\times 10^{-4}$   &  -                \\\hline
 $\Upsilon(2S)\rightarrow \eta_b+\gamma$ & $(3.90\pm 1.49)\times 10^{-4}$ &
  $(3.90\pm 1.49)\times 10^{-4}$ &
  $(3.90\pm 1.49)\times 10^{-4}$      & $(3.9\pm1.1^{+1.1}_{-0.9})\times 10^{-4}$ \cite{:2009pz}  \\\hline
 $\Upsilon(3S)\rightarrow \eta_b+\gamma$ & $(1.87\pm 0.71)\times 10^{-4}$ &
  $(1.68\pm 0.72)\times 10^{-4}$ &
  $(1.05\pm 0.40)\times 10^{-5}$   & $(4.8\pm 0.5\pm 0.6)\times 10^{-4}$  \cite{:2008vj} \\
       &     &         &             & $(7.1\pm 1.8\pm 1.1)\times 10^{-4}$  \cite{Bonvicini:2009hs} \\\hline
 $\Upsilon(4S)\rightarrow \eta_b+\gamma$ & $(8.81\pm 3.32)\times 10^{-8}$ &
  $(7.82\pm 3.35)\times 10^{-8}$ &  $(2.25\pm 0.88)\times 10^{-10}$& -                \\\hline
 $\Upsilon(5S)\rightarrow \eta_b+\gamma$ & $(1.17\pm 0.43)\times 10^{-8}$ &
  $(1.02\pm 0.45)\times 10^{-8}$    & $(1.57\pm 0.52)\times 10^{-12}$ & -                \\
\end{tabular}
\end{ruledtabular}\label{tab:etab2}
\end{table}


\section{Conclusion}

The LFQM has been successful in phenomenological applications. It is
believed that it could be a reasonable model for dealing with the
hadronic transitions where the non-perturbative QCD effects
dominate. However, it seems that the wave function adopted in the
previous literature has to be modified. As we study the decay
constant of $\Upsilon(nS)$, we find that there exists a sharp
contradiction between the theoretical prediction and data as long as
the traditional harmonic oscillator wave functions were employed. Namely,
the larger $n$ is, the larger the predicted decay constant would be. It is
obviously contradict to the physics picture that for higher radially
excited states, the wave function at origin should be smaller than
the lower ones. But the old wave functions would result in an inverse
tendency. If enforcing all the decay constants of $\Upsilon(nS)$ to
be fitted to the data in terms of the traditional wave functions, the
orthogonality among all the $nS$ states must be abandoned, but it is
not acceptable according to the basic principle of quantum
mechanics.

Thus we modify the wave functions of the radial excited states based
on the common principles. Namely, we keep the orthogonality among
the wave functions and their proper normalization. Moreover, we
require the wave functions $\varphi_M(r)$ at origin $r=0$ to be
consistent with the data, i.e. the decay constants for higher $n$
must be smaller than that of the lower states. Concretely, we modify
the exponential function in the wave functions by demanding the
power not to universal for all $n$'s, but be dependent on $n$.
Concretely we add a numerical factor $g_n$ into $\exp(g_n{-{\bf
p}^2\over 2\beta^2})$ and by fitting the data of the decay constants
of $\Upsilon(nS)$ we obtain a series of the numbers of $g_n$. Within
a reasonable error range, we approximate $g_n$ as $g(n)=n^\delta$
and calculate the value for $\delta$. It is an alternative way which
is different from that adopted in Ref. \cite{Choi:2007se}, to fix
the parameter.

With the modified wave functions of $\Upsilon(nS)$, we calculate
the branching ratios of $\Upsilon(nS)\rightarrow\eta_b+\gamma$ in
the LFQM.  First by fitting the well-measured central value of
$\mathcal{B}(\Upsilon(2S)\to \eta_b+\gamma)$ \cite{:2009pz}, we
obtain the parameter $\beta_{\eta_b}$. By the effective heavy
quark theory, in heavy quark limit the spin singlet and triplet of
$b\bar b$ system should degenerate, namely the parameters of
$\beta_{\Upsilon(1S)}$ and $\beta_{\eta_b}$ should be very close.
Our numerical result confirms this requirement.

Then we estimate the other $\Upsilon(nS)\to \eta_b+\gamma$. The
order of magnitudes of our numerical results is consistent with
data. Even though the predicted branching ratios still do not
precisely coincide with the data, the result is much improved.
The branching ratios of processes $\Upsilon(4S)\to \eta_b+\gamma$
and $\Upsilon(5S)\to \eta_b+\gamma$ are predicted to be at the order
of $10^{-8}$. They are difficult to be measured in the future as
long as there is no new physical mechanism to greatly enhance them.



By studying the radiative decay of $\Upsilon(nS)\to \eta_b+\gamma$,
we can learn much about the hadronic structure of $\eta_b$. Even
though much effort has been made to explore the spin singlet
$\eta_b$, in particle data group (PDG) of 2008, $\eta_b$ was still omitted from
the summary table \cite{PDG08}. In fact, determination of the mass
of $\eta_b$ is made via the radiative decays of $\Upsilon(nS)\to
\eta_b+\gamma$ \cite{:2008vj}, and the recent data show
$m_{\eta_b}=9388.9^{+3.1}_{-2.3}(stat)\pm 2.7(syst)$ MeV by the
$\Upsilon(3S)$ data and $m_{\eta_b}=9394.2^{+4.8}_{-4.9}(stat)\pm
2.0(syst)$ MeV by the $\Upsilon(2S)$ data \cite{:2009pz}. Penin
\cite{Penin:2009wf} reviewed the progress for determining the mass
of $\eta_b$ and indicated that the accurate theoretical prediction
of $m_{\eta_b}$ would be a great challenge. Indeed, determining the
wave function of $\eta_b$ would be even more challenging. We
carefully study the transition rates of the radiative decays which
would help to extract information about $m_{\eta_b}$. The transition
rate of $\Upsilon(1S)\to\eta_b+\gamma$ is very sensitive to the mass
splitting $\Delta M=m_{\Upsilon(1S)}-m_{\eta_b}$ due to the phase
space constraint, thus an accurate measurement of the radiative
decay may be more useful to learn the spin dependence of the
bottominia.

\section*{Acknowledgments}
This project is supported by the National Natural Science
Foundation of China (NSFC) under Contracts Nos. 10705001, 10705015
and 10775073; the Foundation for the Author of National Excellent
Doctoral Dissertation of P.R. China (FANEDD) under Contracts No.
200924; the Doctoral Program Foundation of Institutions of Higher
Education of P.R. China under Grant No. 20090211120029; the
Special Grant for the Ph.D. program of Ministry of Eduction of
P.R. China; the Program for New Century Excellent Talents in
University (NCET) by Ministry of Education of P.R. China; the Fundamental Research Funds for the Central Universities; the
Special Grant for New Faculty from Tianjin University.

\section*{Appendix}
\subsection{The radial wave functions}

The traditional wave functions $\phi$ in configuration space from harmonic oscillator \cite{Faiman:1968js}
are
\begin{eqnarray}\label{app4}
\varphi^{1S}(r)&=&\Big(\frac{\beta^2}{\pi}\Big)^{3/4}{\exp}\Big(-\frac{1}{2}\beta^2\mathbf{r}^2\Big),\nonumber\\
\varphi^{2S}(r)&=&\Big(\frac{\beta^2}{\pi}\Big)^{3/4}{\exp}\Big(-\frac{1}{2}\beta^2\mathbf{r}^2\Big)\frac{1}{\sqrt{6}}
\Big(3-2\beta^2\mathbf{r}^2\Big),
\nonumber\\
\varphi^{3S}(r)&=&\Big(\frac{\beta^2}{\pi}\Big)^{3/4}{\exp}\Big(-\frac{1}{2}\beta^2\mathbf{r}^2\Big)
\sqrt{\frac{2}{15}}
\Big(\frac{15}{4}-5\beta^2\mathbf{r}^2+\beta^4\mathbf{r}^4\Big),\nonumber \\
\varphi^{4S}(r)&=&\Big(\frac{\beta^2}{\pi}\Big)^{3/4}{\rm
exp}\Big(-\frac{1}{2}\mathbf{r}^2\beta^2\Big)\frac{1}{{12\sqrt{35}}}
\Big(-105+210\mathbf{r}^2{\beta^2}-84\mathbf{r}^4{\beta^4}+8\mathbf{r}^6{\beta^6}\Big),
\nonumber\\
\varphi^{5S}(r)&=&\Big(\frac{\beta^2}{\pi}\Big)^{3/4}{\rm
exp}\Big(-\frac{1}{2}\mathbf{r}^2\beta^2\Big)\frac{1}{{72\sqrt{70}}}
\Big(945-2520{\beta^2}\mathbf{r}^2+1512{\beta^4}\mathbf{r}^4
-288{\beta^6}\mathbf{r}^6+ 16{\beta^8}\mathbf{r}^8\Big).
 \end{eqnarray}
and their Fourier transformation are
\begin{eqnarray}\label{app5}
\psi^{1S}(\mathbf{p}^2)&=&\Big(\frac{1}{\beta^2\pi}\Big)^{3/4}{\exp}\Big(-\frac{1}{2}\frac{\mathbf{p}^2}{\beta^2}\Big),\nonumber\\
\psi^{2S}(\mathbf{p}^2)&=&\Big(\frac{1}{\beta^2\pi}\Big)^{3/4}{\exp}\Big(-\frac{1}{2}\frac{\mathbf{p}^2}{\beta^2}\Big)\frac{1}{\sqrt{6}}
\Big(3-2\frac{\mathbf{p}^2}{\beta^2}\Big),
\nonumber\\
 \psi^{3S}(\mathbf{p}^2)&=&\Big(\frac{1}{\beta^2\pi}\Big)^{3/4}{\exp}\Big(-\frac{1}{2}\frac{\mathbf{p}^2}{\beta^2}\Big)
\sqrt{\frac{2}{15}}
\Big(\frac{15}{4}-5\frac{\mathbf{p}^2}{\beta^2}+\frac{\mathbf{p}^4}{\beta^4}\Big),\nonumber \\
 \psi^{4S}(\mathbf{p}^2)&=&\Big(\frac{1}{\beta^2\pi}\Big)^{3/4}{\rm
exp}\Big(-\frac{1}{2}\frac{\mathbf{p}^2}{\beta^2}\Big)\frac{1}{{12\sqrt{35}}}
\Big(-105+210\frac{\mathbf{p}^2}{\beta^2}-84\frac{\mathbf{p}^4}{\beta^4}+8\frac{\mathbf{p}^6}{\beta^6}\Big),
\nonumber\\
\psi^{5S}(\mathbf{p}^2)&=&\Big(\frac{1}{\beta^2\pi}\Big)^{3/4}{\rm
exp}\Big(-\frac{1}{2}\frac{\mathbf{p}^2}{\beta^2}\Big)\frac{1}{{72\sqrt{70}}}
\Big(945-2520\frac{\mathbf{p}^2}{\beta^2}+1512\frac{\mathbf{p}^4}{\beta^4}
-288\frac{\mathbf{p}^6}{\beta^6}+
16\frac{\mathbf{p}^8}{\beta^8}\Big).
 \end{eqnarray}

The modified  wave functions $\varphi_M$ in configuration space are
defined
\begin{eqnarray}\label{app4}
\varphi_{\rm_M}^{1S}(r)&&=\Big(\frac{\beta^2}{\pi}\Big)^{3/4}{\exp}\Big(-\frac{1}{2}\beta^2\mathbf{r}^2\Big),\nonumber\\
\varphi_{\rm_M}^{2S}(r)&&=\Big(\frac{\beta^2}{\pi}\Big)^{3/4}{\exp}\Big(-\frac{1}{2\times{2}^\delta}\beta^2\mathbf{r}^2\Big)
\Big(a_2 - b_2\beta^2\mathbf{r}^2\Big),
\nonumber\\
\varphi_{\rm_M}^{3S}(r)&&=\Big(\frac{\beta^2}{\pi}\Big)^{3/4}{\exp}\Big(-\frac{1}{2\times{3}^\delta}\beta^2\mathbf{r}^2\Big)
\Big(a_3 - b_3\beta^2\mathbf{r}^2+c_3\beta^4\mathbf{r}^4\Big),\nonumber\\
\varphi_{\rm_M}^{4S}(r)&&=\Big(\frac{\beta^2}{\pi}\Big)^{3/4}{\rm
exp}\Big(-\frac{1}{2\times{4}^\delta}\mathbf{r}^2\beta^2\Big)
\Big(-a_4 +
b_4\mathbf{r}^2{\beta^2}-c_4\mathbf{r}^4{\beta^4}+d_4\mathbf{r}^6{\beta^6}\Big),
\nonumber\\
\varphi_{\rm_M}^{5S}(r)&&=\Big(\frac{\beta^2}{\pi}\Big)^{3/4}{\rm
exp}\Big(-\frac{1}{2\times{5}^\delta}\mathbf{r}^2\beta^2\Big)
\Big(a_5 - b_5{\beta^2}\mathbf{r}^2+c_5{\beta^4}\mathbf{r}^4
-d_5{\beta^6}\mathbf{r}^6 +e_5{\beta^8}\mathbf{r}^8\Big)
 \end{eqnarray}
with coefficients, which are irrational numbers and are kept five
digits after the decimal point
\begin{eqnarray*}
\begin{array}{|c|c|c|c|c|c|}\toprule[1pt]
n&a_n&b_n&c_n&d_n&e_n\\\hline
2& 0.72817& 0.40857&-&-&-\\
3& 0.62920& 0.54138&0.06712&-&-\\
4&0.57834&0.61887&0.12838&0.00614&-\\
5&0.54747&0.67621&0.18332&0.01558&0.00038
\\\bottomrule[1pt]
\end{array}\,.
\end{eqnarray*}

The corresponding modified wave functions in momentum space are
\begin{eqnarray}\label{app4}
\psi_{\rm_M}^{1S}(\mathbf{p}^2)&&=\Big(\frac{1}{\beta^2\pi}\Big)^{3/4}{\exp}\Big(-\frac{1}{2}\frac{\mathbf{p}^2}{\beta^2}\Big),\nonumber\\
\psi_{\rm_M}^{2S}(\mathbf{p}^2)&&=\Big(\frac{1}{\beta^2\pi}\Big)^{3/4}{\exp}\Big(-\frac{{2}^\delta}{2}\frac{\mathbf{p}^2}{\beta^2}\Big)
\Big(a^\prime_2 -b^\prime_2\frac{\mathbf{p}^2}{\beta^2}\Big),\nonumber\\
\psi_{\rm_M}^{3S}(\mathbf{p}^2)&&=\Big(\frac{1}{\beta^2\pi}\Big)^{3/4}{\exp}\Big(-\frac{{3}^\delta}{2}\frac{\mathbf{p}^2}{\beta^2}\Big)
\Big(a^\prime_3 - b^\prime_3\frac{\mathbf{p}^2}{\beta^2}+c^\prime_3\frac{\mathbf{p}^4}{\beta^4}\Big),\nonumber\\
\psi_{\rm_M}^{4S}(\mathbf{p}^2)&&=\Big(\frac{1}{\beta^2\pi}\Big)^{3/4}{\rm
exp}\Big(-\frac{{4}^\delta\mathbf{p}^2}{2\beta^2}\Big)
\Big(-a^\prime_4+ b^\prime_4\frac{\mathbf{p}^2}{\beta^2}
-c^\prime_4\frac{\mathbf{p}^4}{\beta^4}+d^\prime_4\frac{\mathbf{p}^6}{\beta^6}\Big),\nonumber\\
\psi_{\rm_M}^{5S}(\mathbf{p}^2)&&=\Big(\frac{1}{\beta^2\pi}\Big)^{3/4}{\rm
exp}\Big(-\frac{{5}^\delta}{2}\frac{\mathbf{p}^2}{\beta^2}\Big)
\Big(a^\prime_5
-b^\prime_5\frac{\mathbf{p}^2}{\beta^2}+c^\prime_5\frac{\mathbf{p}^4}{\beta^4}
-d^\prime_5\frac{\mathbf{p}^6}{\beta^6}+e^\prime_5\frac{\mathbf{p}^8}{\beta^8}\Big)
 \end{eqnarray}
with coefficients
\begin{eqnarray*}
\begin{array}{|c|c|c|c|c|c|}\toprule[1pt]
n&a^\prime_n&b^\prime_n&c^\prime_n&d^\prime_n&e^\prime_n\\\hline
2&1.88684& 1.54943&-&-&-\\
3&2.53764& 5.67431&1.85652&-&-\\
4&3.1439&12.58984&10.05113&1.88915&-\\
5&3.67493&22.58205&31.06666&13.51792&1.70476
\\\bottomrule[1pt]
\end{array}\,.
\end{eqnarray*}

\subsection{Some notations in LFQM}
The incoming (outgoing) meson in Fig. \ref{fig:LFQM} has the
momentum ${P}^({'}^)={p_1}^({'}^)+p_2$ where ${p_1}^({'}^)$ and
$p_2$ are the momenta of the off-shell quark and antiquark and
\begin{eqnarray}\label{20}
p^+_1&=&x_1 P^+,\qquad p^+_2 = x_2 P^+,
\nonumber\\
{ p}_{1\perp}&=& x_1{ P}_\perp + { k}_\perp,\qquad { p}_{2\perp}= x_2{
P}_\perp - { k}_\perp,
\nonumber\\
p'^+_1&=&x_1 P^+,\qquad p'^+_2 = x_2 P^+,
\nonumber\\
{ p'}_{1\perp}&=& x_1{ P'}_\perp + { k'}_\perp, \qquad{ p'}_{2\perp}=
x_2{ P'}_\perp - { k'}_\perp\nonumber
\end{eqnarray}
with $x_1+x_2=1$, where $x_i$ and $k_\perp(k'_\perp)$ are internal
variables. $M_0$ and $\tilde {M_0}$ are defined
\begin{eqnarray}\label{app2}
&&M_0^2=\frac{k^2_\perp+m^2_1}{x_1}+\frac{k^2_\perp+m^2_2}{x_2},\nonumber\\&&
\tilde {M_0}=\sqrt{M_0^2-(m_1-m_2)^2}.\nonumber
 \end{eqnarray}

The  wave functions $\phi_M$ are transformed into
\begin{eqnarray}\label{app4}
\phi_{\rm_M}(1S)&&=4\Big(\frac{\pi}{\beta^2}\Big)^{3/4}\sqrt{\frac{\partial
k_z}{\partial x}}{\exp}\Big(-\frac{k^2_z+k^2_\perp}{2\beta^2}\Big),\nonumber\\
\phi_{\rm_M}(2S)&&=4\Big(\frac{\pi}{\beta^2}\Big)^{3/4}\sqrt{\frac{\partial
k_z}{\partial
x}}{\exp}\Big(-\frac{{2}^\delta}{2}\frac{k^2_z+k^2_\perp}{\beta^2}\Big)
\Big(a_2^\prime -b_2^\prime\frac{k^2_z+k^2_\perp}{\beta^2}\Big),
\nonumber\\
\phi_{\rm_M}(3S)&&=4\Big(\frac{\pi}{\beta^2}\Big)^{3/4}\sqrt{\frac{\partial
k_z}{\partial
x}}{\exp}\Big(-\frac{{3}^\delta}{2}\frac{k^2_z+k^2_\perp}{\beta^2}\Big)
\Big(a_3^\prime -
b_3^\prime\frac{k^2_z+k^2_\perp}{\beta^2}+c_3^\prime\frac{(k^2_z+k^2_\perp)^2}{\beta^4}\Big),\nonumber\\
\phi_{\rm_M}(4S)&&=4\Big(\frac{\pi}{\beta^2}\Big)^{3/4}\sqrt{\frac{\partial
k_z}{\partial x}}{\rm
exp}\Big(-\frac{{4}^\delta}{2}\frac{k^2_z+k^2_\perp}{\beta^2}\Big)
\Big(-a_4^\prime + b_4^\prime\frac{k^2_z+k^2_\perp}{\beta^2}
-c_4^\prime\frac{(k^2_z+k^2_\perp)^2}{\beta^4}+d_4^\prime\frac{(k^2_z+k^2_\perp)^3}{\beta^6}\Big),
\nonumber\\
\phi_{\rm_M}(5S)&&=4\Big(\frac{\pi}{\beta^2}\Big)^{3/4}\sqrt{\frac{\partial
k_z}{\partial x}}{\rm
exp}\Big(-\frac{{5}^\delta}{2}\frac{k^2_z+k^2_\perp}{\beta^2}\Big)
\Big(a_5^\prime -
b_5^\prime\frac{k^2_z+k^2_\perp}{\beta^2}+c_5^\prime\frac{(k^2_z+k^2_\perp)^2}{\beta^4}
-d_5^\prime\frac{(k^2_z+k^2_\perp)^3}{\beta^6}+e_5^\prime\frac{(k^2_z+k^2_\perp)^4}{\beta^8}\Big).\nonumber\\
 \end{eqnarray}

More information can be found in Ref. \cite{Cheng:2003sm}.

\end{document}